\documentclass[final]{aipproc}
\layoutstyle{6x9}
\usepackage{graphicx,amsmath,wrapfig,epsfig}
\begin{document}
\title{Selected topics \\ on parton distribution functions}
\classification{13.60.Hb, 13.60.-r, 24.85.+p, 13.88.+e, 13.87.-a}
\keywords{Quark, gluon, parton, distribution, QCD, nuclear effect, 
          tensor, jet, CDF anomaly}
\author{M. Hirai}{
address={Department of Physics, Faculty of Science and Technology,
         Tokyo University of Science \\ 
         2641, Yamazaki, Noda, Chiba, 278-8510, Japan}}
\author{H. Kawamura}{
  address={KEK Theory Center, Institute of Particle and Nuclear Studies, KEK \\
           1-1, Ooho, Tsukuba, Ibaraki, 305-0801, Japan}}
\author{S. Kumano}{
  address={KEK Theory Center, Institute of Particle and Nuclear Studies, KEK \\
           1-1, Ooho, Tsukuba, Ibaraki, 305-0801, Japan}
 ,altaddress={J-PARC Branch, KEK Theory Center,
              Institute of Particle and Nuclear Studies, KEK \\
              and
              Theory Group, Particle and Nuclear Physics Division, 
              J-PARC Center \\
              203-1, Shirakata, Tokai, Ibaraki, 319-1106, Japan}}
\author{K. Saito}{
address={Department of Physics, Faculty of Science and Technology,
         Tokyo University of Science \\ 
         2641, Yamazaki, Noda, Chiba, 278-8510, Japan}}

\begin{abstract}
We report recent studies on structure functions of the nucleon
and nuclei. First, clustering effects are investigated
in the structure function $F_2$ of $^9$Be for explaining an unusual nuclear
correction found in a JLab experiment. We propose that high densities
created by formation of clustering structure like $2 \alpha$+neutron
in $^9$Be is the origin of the unexpected JLab result by using 
the antisymmetrized molecular dynamics (AMD).
There is an approved proposal at JLab to investigate 
the structure functions of light nuclei including the cluster structure, 
so that much details will become clear in a few years.
Second, tensor-polarized quark and antiquark distributions are obtained
by analyzing HERMES measurements on the structure function $b_1$
for the deuteron. The result suggests a finite tensor polarization
for antiquark distributions, which is an interesting topic for 
further theoretical and experimental investigations. An experimental
proposal exists at JLab for measuring $b_1$ of the deuteron 
as a new tensor-structure study in 2010's.
Furthermore, the antiquark tensor polarization could be measured
by polarized deuteron Drell-Yan processes at hadron facilities
such as J-PARC and GSI-FAIR. 
Third, the recent CDF dijet anomaly is investigated within
the standard model by considering possible modifications 
of the strange-quark distribution. We find that the shape of
a dijet-mass spectrum changes depending on the strange-quark
distribution. It indicates that the CDF excess could be partially
explained as a PDF effect, particularly by the strangeness
in the nucleon, within the standard model if the excess at 
$m_{jj} \approx$140 GeV is not a sharp peak.
\end{abstract}

\maketitle

\section{Introduction}
\vspace{-0.15cm}

Studies of parton distribution functions (PDFs) are important 
for investigating internal structure of hadrons and
for calculating precise cross sections in hadron reactions.
The precise information is necessary for finding any
new physics beyond the standard model and new phenomena
in hadron and nuclear physics. Recently, there are interesting
findings and progresses on high-energy hadron reactions. 
We explain the following selected topics in this report:
\vspace{-0.15cm}
\begin{itemize}
\item[1.] a possible clustering effect in the structure function $F_2$
          of the $^9$Be nucleus \cite{jlab-09,hksw-cluster},
\item[2.] tensor structure in quark and gluon degrees of freedom
          by the structure function $b_1$ 
          \cite{hermes-b1,sk-b1-analysis},
\item[3.] test of CDF dijet anomaly within the standard model
          \cite{cdf-w2j-2011,kek-2011}.
\end{itemize}
\vspace{-0.15cm}
We explain these topics by the following sections.

\vfill\eject

\section{Nuclear clustering effects}
\vspace{-0.15cm}

\begin{wrapfigure}{r}{0.45\textwidth}
   \vspace{0.0cm}
   \begin{center}
       \epsfig{file=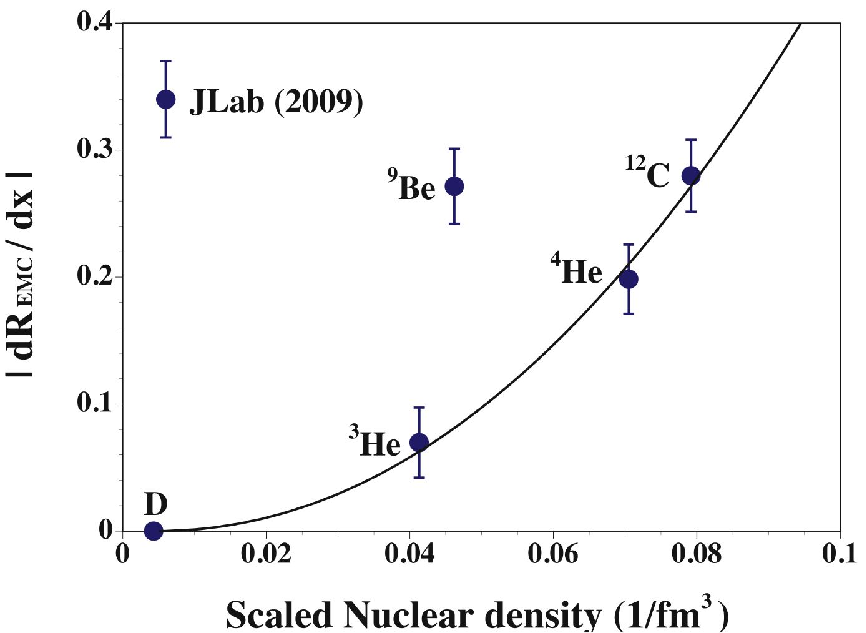,width=0.40\textwidth} \\
   \end{center}
   \vspace{-0.2cm}
{\footnotesize \hspace{0.75cm}
 {\bf FIGURE 1.} Nuclear modifications 

\noindent
\hspace{0.75cm} 
  $| d(F_2^A/F_2^D)/dx |$ for light nuclei.
 }
\vspace{0.2cm}
\end{wrapfigure}

Nuclear modifications of the structure function $F_2$ 
were recently measured at the Thomas Jefferson National
Accelerator Facility (JLab) for light nuclei.
By taking the $x$-slope of the ratio $F_2^A/F_2^D$,
namely $|d (F_2^A/F_2^D) /dx|$, they observed 
anonymously large nuclear effect in $^9$Be as shown
in Fig.1. It is too large to be expected from 
an average nuclear density.

It is known that clustering structure is expected to exist 
for light nuclei around the mass number $A=10$ according to
theoretical nuclear models such as antisymmetrized (or fermionic)
molecular dynamics (AMD or FMD). 
We consider that an anomalous nuclear modification of 
the $^9$Be structure function could be related to
the nuclear cluster structure. We investigated a possible clustering
effect by using the AMD and shell-model wave functions
for the $^9$Be nucleus. We show a possibility to explain
the anomalous JLab result by cluster formation \cite{hksw-cluster}.

Nuclear structure functions are described by
a convolution picture \cite{sumemc,ek03}:
\begin{align}
\! \! \! \! \!
F_{2}^A (x, Q^2) & = \int_x^A dy \, f(y) \, F_{2}^N (x/y, Q^2) , \ \ 
f(y)  =  \frac{1}{A} \int d^3 p_N
     \, y \, \delta \left( y - \frac{p_N \cdot q}{M_N \nu} \right) 
     | \phi (\vec p_N) |^2 ,
\label{eqn:w-convolution}
\end{align}
where $y$ is the momentum fraction 
$ y   =  M_A \, p_N \cdot q /(M_N \, p_A \cdot q) $.
The nucleon momentum distribution $\phi (\vec p_N)$ is calculated
in two models. One is the simple shell model, and the other
is the AMD in order to illustrate clustering effects
by differences between the two-model results.
An advantage of the AMD is that there is no assumption 
on nuclear structure, namely shell or cluster-like configuration.
The spacial density distributions are shown in 
Fig.2 for $^4$He and Fig.3 for $^9$Be.
The density distributions are the same in both shell-model
and AMD for $^4$He. However, the AMD density of $^9$Be 
in Fig.3 is much different from a monotonic shell-model 
density similar to the one in Fig.2. It is obvious that $^9$Be 
has two $\alpha$ clusters with neutron clouds.

\vspace{0.5cm}
\begin{figure}[h]
   \hspace{0.2cm}
   \includegraphics[width=0.40\textwidth]{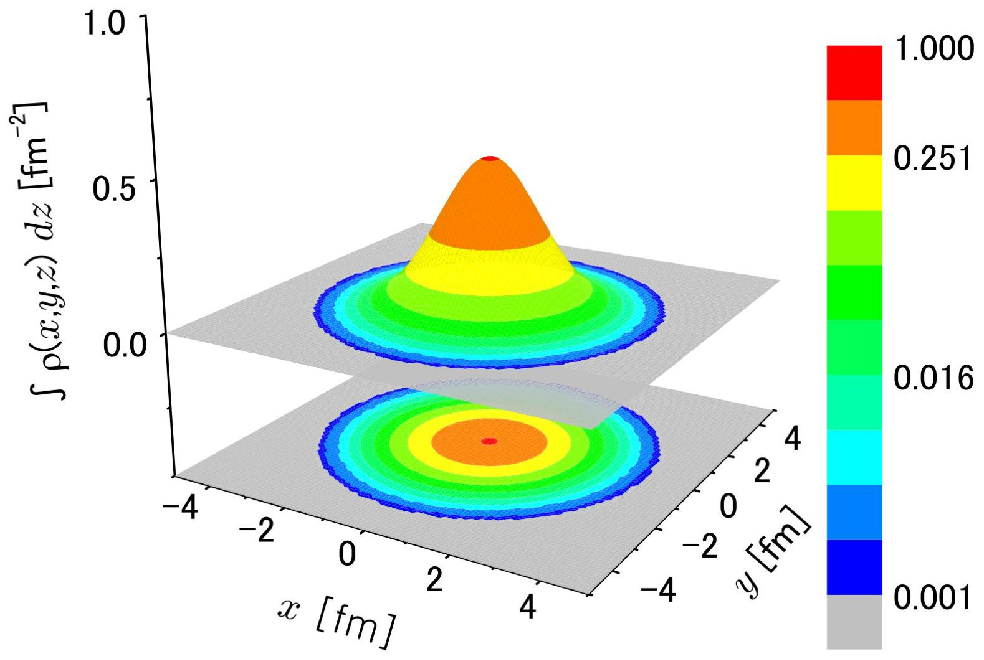}
   \hspace{1.5cm}
   \includegraphics[width=0.40\textwidth]{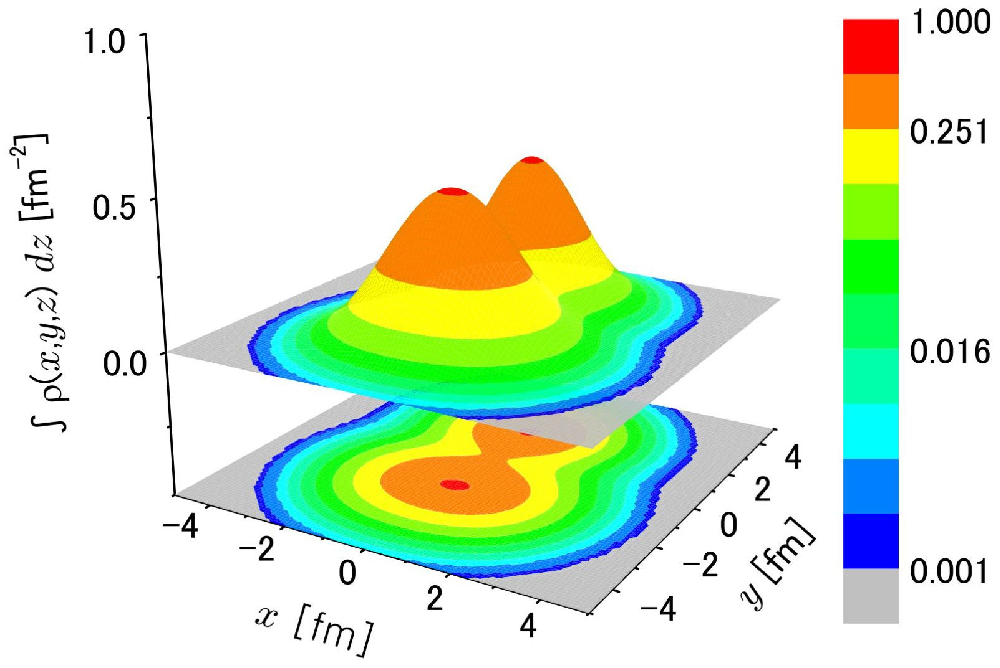}
  $ \ \ \ \ \ \ \ \ \ \ \ $
\label{fig:b1-motivation}
\end{figure}
\vspace{0.0cm}
\noindent
{\footnotesize 
 {\bf FIGURE 2.} Density distribution of $^4$He
                 \cite{hksw-cluster}.
 \hspace{0.6cm}
 {\bf FIGURE 3.} Density distribution of $^9$Be by AMD
                 \cite{hksw-cluster}.
 }

\vfill\eject

\begin{wrapfigure}{r}{0.50\textwidth}
   \vspace{0.0cm}
   \begin{center}
       \epsfig{file=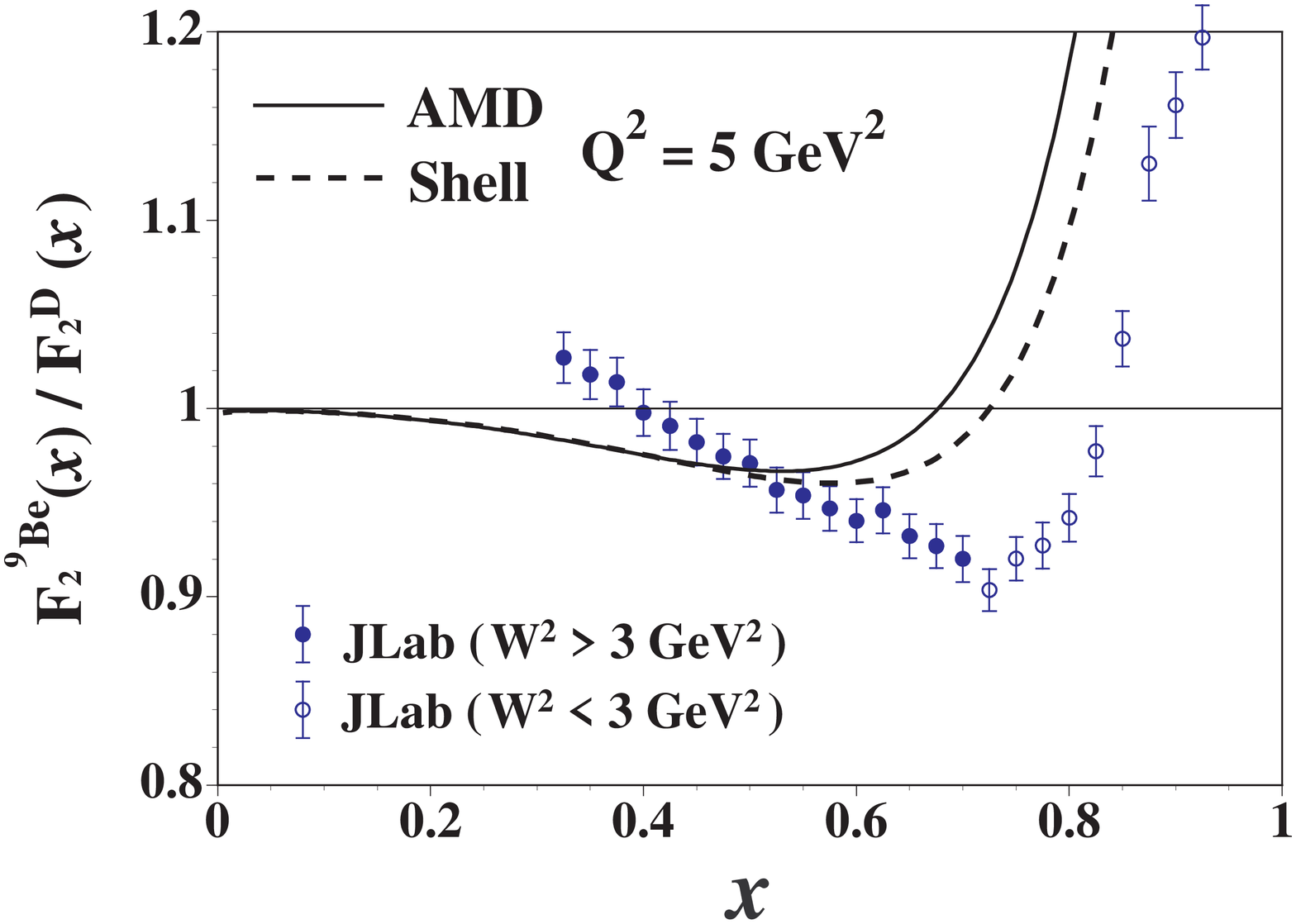,width=0.45\textwidth} \\
   \end{center}
   \vspace{-0.2cm}
{\footnotesize \hspace{1.30cm}
 {\bf FIGURE 4.} $F_2^{^9 Be} / F_2^D$ by shell and 

\vspace{-0.05cm}
\noindent
\hspace{1.50cm} 
  AMD models \cite{hksw-cluster}.
 }
\vspace{0.4cm}
   \begin{center}
       \epsfig{file=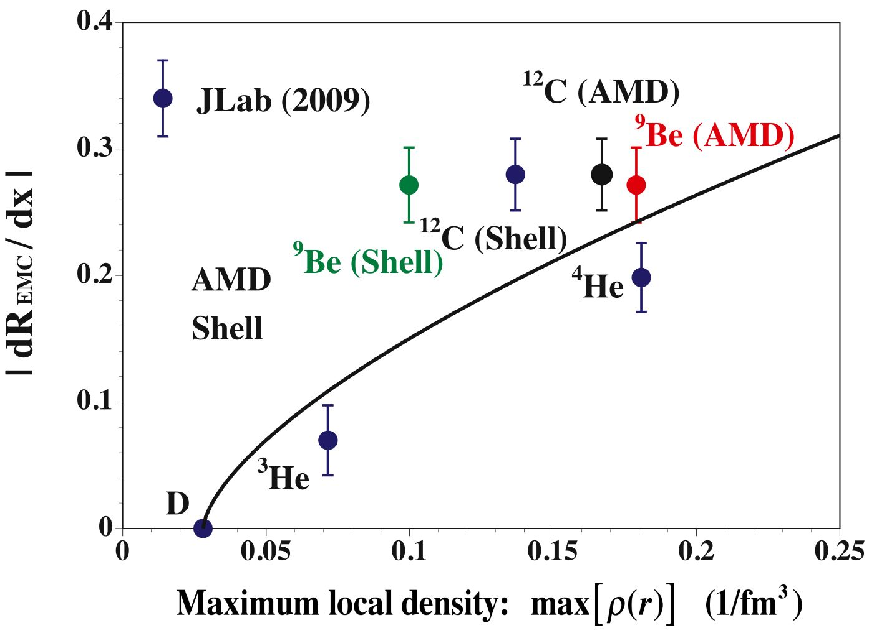,width=0.43\textwidth} \\
   \end{center}
   \vspace{-0.2cm}
{\footnotesize \hspace{0.70cm}
 {\bf FIGURE 5.} Nuclear modification slopes 
 
\noindent
\hspace{0.90cm}
                 shown by maximum local densities \cite{hksw-cluster}.
 }
\vspace{0.4cm}
\end{wrapfigure}

Calculated nuclear modifications are shown in Fig.4.
Although the nuclear density distributions are much different
in the shell and AMD models, the clustering effects,
namely the differences between the solid and dashed curves,
are not very large in Fig.4 due to angular average in 
calculating the convolution integral.
The theoretical curves do not agree quite well with the experimental data,
which is partly because short-range-correlation effects are not 
taken into account in the nucleon momentum distribution
and the separation energy.
The small difference between the shell and AMD results 
suggests that the mean conventional mechanism
does not solve the anomalously large nuclear modification.

Next, we consider that the high-densities created by the formation
of clusters in Fig. 3 could be the origin of the large JLab modification.
Obviously, the density peaks in the two clusters are much larger than
shell-model densities. We calculated maximum local densities
in both shell and AMD models. The JLab data are shown by
the calculated maximum local densities in Fig.5.
The curve is the interpolation of shell-model data points
except for $^9$Be. If the $^9$Be data point is shown by
the density of the shell model, it significantly deviates from the curve.
However, if it is plotted by the AMD density, it agrees with the curve.
This fact suggests that 
{\it the JLab result should be caused by
the high densities created by the cluster formation in $^9$Be}
\cite{hksw-cluster}.

We consider that the nuclear structure functions consist 
of the mean conventional part and the remaining one depending 
on the maximum local density:
\vspace{-0.05cm}
\begin{equation}
F_2^A = \text{(mean part)} 
       +\text{(part created by large densities
               due to cluster formation)}.
\end{equation}
\vspace{-0.05cm}\hspace{-0.15cm}
\noindent
The first part is calculated by the convolution integral of
Eq.(\ref{eqn:w-convolution}) and the second part seems to 
play a crucial role in explaining the JLab anomaly.
The physics mechanism for the second term is likely to be the modification
of internal nucleon structure caused by the cluster formation.

Our studies in Ref. \cite{hksw-cluster} stimulated JLab experimentalists
to investigate possible cluster structure in 
an approved experimental proposal to measure the structure functions
of light nuclei \cite{jlab-cluster-exp}
including $^6$Li, $^7$Li, $^{10}$B, and $^{11}$B in addition to $^9$Be.
Therefore, much details
should become clear in a few years by precise JLab measurements
after 12 GeV upgrade.

\vfill\eject

\section{Tensor structure function}
\vspace{-0.2cm}

Tensor structure of the deuteron has been investigated
for a long time by hadron degrees of freedom as explained in standard
nuclear-physics textbooks. However, the time has come to describe
it in terms of quark and gluon degrees of freedom.  
In particular, the first measurement of the tensor structure
function $b_1$ has been reported by the HERMES collaboration
\cite{hermes-b1}.
As illustrated in Fig.6, $b_1$ vanishes if constituents are
in the $S$-wave. In the ``standard model" of the deuteron
with $D$-state admixture, a finite $b_1$ is obtained, for example,
in the convolution picture for the structure function.
If experimental measurements are different from this standard-model 
predication, it should provide opportunities for interesting new
aspects in the tensor structure.

\vspace{0.4cm}
\begin{figure}[h!]
   \includegraphics[width=0.95\textwidth]{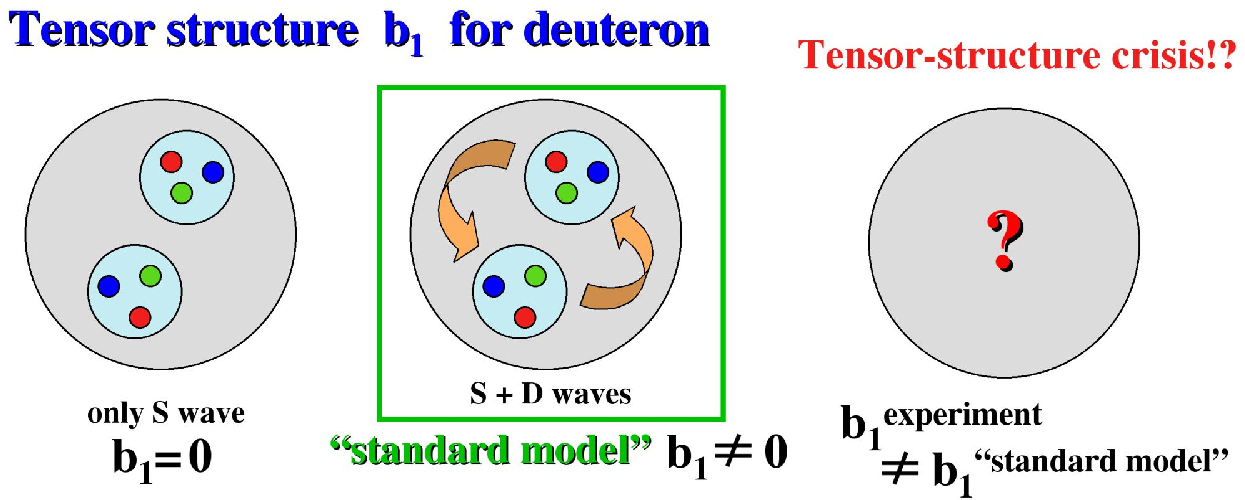}
\label{fig:b1-motivation}
\end{figure}
\vspace{-0.1cm}
\hspace{0.6cm}
\begin{minipage}[c]{0.85\textwidth}
\setlength{\baselineskip}{10pt}
\vspace{0.5cm}
\hspace{2.0cm}
{\footnotesize {\bf FIGURE 6.} Illustration of tensor structure
 in the deuteron.}
\end{minipage}
\vspace{0.8cm}

In this work, tensor polarized quark and antiquark
distribution functions are extracted from the HERMES measurement.
The $b_1$ is expressed by tensor-polarized distributions $\delta_T q$ as
\begin{equation}
b_1 (x) = \frac{1}{2} \sum_i e_i^2 
   \left[ \delta_T q_i(x) 
        + \delta_T \bar q_i(x) \right ] , \ \ \ 
\delta_T q_i (x) = q^0_i (x)
    -\frac{q^{+1}_i (x) +q^{-1}_i (x)}{2}  , 
\end{equation}
where $q_i^\lambda$ indicates an unpolarized-quark distribution
in the hadron spin state $\lambda$.
In analyzing $b_1$ data, the distributions $\delta_T q$ and
 $\delta_T \bar q$ are parametrized as \cite{sk-b1-analysis}
\begin{equation}
\delta_T q_{iv}^D (x) = \delta_T w(x) \, q_{iv}^D (x) , \ 
\delta_T \bar q_i^D (x) 
           = \alpha_{\bar q} \, \delta_T w(x) \, \bar q_i^D (x) , \ 
\delta_T w(x) = a x^b (1-x)^c (x_0-x) ,
\label{eqn:dw(x)-abc}
\end{equation}
by considering that certain fractions of unpolarized distributions
are tensor polarized. Here, flavor-symmetric antiquark distribution
\cite{flavor} are assumed for tensor polarization.
The parameters $\alpha_{\bar q}$, $a$, $b$, and $c$ are determined 
by analyzing the HERMES data with the following two options: 

\vspace{0.15cm} 
$\! \! \! \! \bullet$ Set 1: Tensor-polarized antiquark distributions 
              are terminated ($\alpha_{\bar q} =0$).

\vspace{0.15cm}
$\! \! \! \! \bullet$ Set 2: Finite tensor-polarized antiquark distributions
             are allowed ($\alpha_{\bar q}$ is a parameter).

\vspace{0.15cm}

Analysis results are shown with the HERMES data in Fig.7.
The dashed curve without the antiquark polarization does not
agree with the data in the small-$x$ region; however, the
solid curve of the set-2 analysis agrees well with the data.
Obtained tensor-polarized quark and antiquark distributions
are shown in Fig.8. It is interesting to find a finite 
tensor-polarized antiquark distributions.
It was shown in Ref. \cite{b1-sum} that the first moment of $b_1$
is related to a finite tensor polarization of antiquarks
in Eq. (\ref{eqn:b1x-sum}):
\begin{align}
& \int dx \, b_1 (x) = - \frac{5}{24} \lim_{t \rightarrow 0} t F_Q (t)
\nonumber \\
& \ \ \ \ \ \ \ \ \ \ \ \ \ \ \ \ \ \ \ \ \ \ 
  + \frac{1}{18} \int dx \, [ \, 8 \delta_T \bar u (x) 
  + 2 \delta_T \bar d(x)  +\delta_T s (x) + \delta_T \bar s(x) \, ] ,
\label{eqn:b1x-sum}   \\            
&  \int \frac{dx}{x} \, [F_2^p (x) - F_2^n (x) ] = 
   \frac{1}{3} 
  +\frac{2}{3} \int dx [ \bar u(x) - \bar d(x) ] ,
\label{eqn:gottfried-sum}               
\end{align}
where $F_Q (t)$ is the quadrupole form factor.
This $b_1$ sum rule is
very similar to the Gottfried sum, which indicates
a finite $\bar u-\bar d$ by the deviation from 1/3
in Eq. (\ref{eqn:gottfried-sum}).
In the analysis of set-2, the sum is $\int dx \, b_1 (x) =0.0058$
\cite{sk-b1-analysis}.
According to Eq. (\ref{eqn:b1x-sum}), it should be related to
a finite antiquark tensor polarization, which is interesting
for further theoretical investigations on a possible 
mechanisms to create it.

\vspace{0.0cm}
\begin{figure}[t]
\begin{minipage}[c]{\textwidth}
\vspace{0.3cm}
   \includegraphics[width=0.46\textwidth]{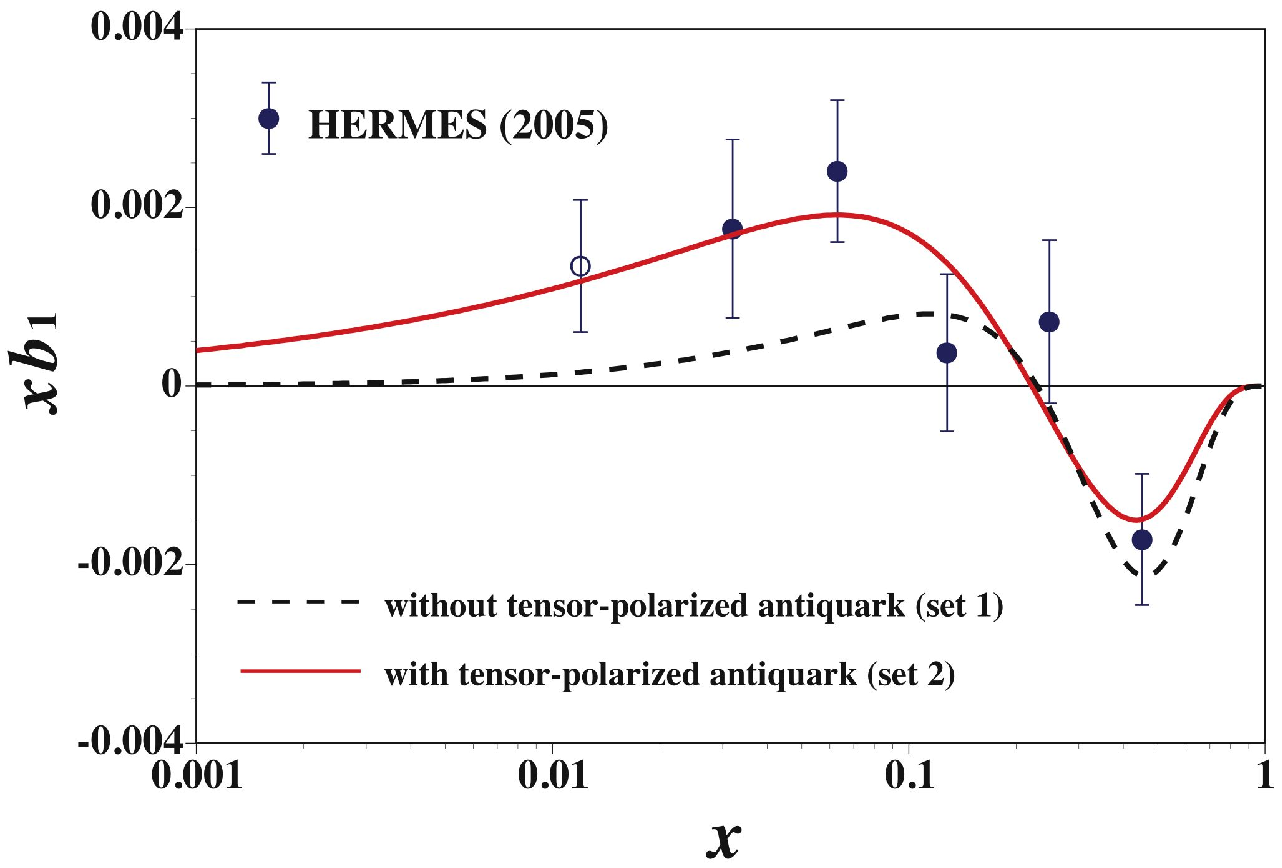}
   \hspace{0.6cm}
   \includegraphics[width=0.46\textwidth]{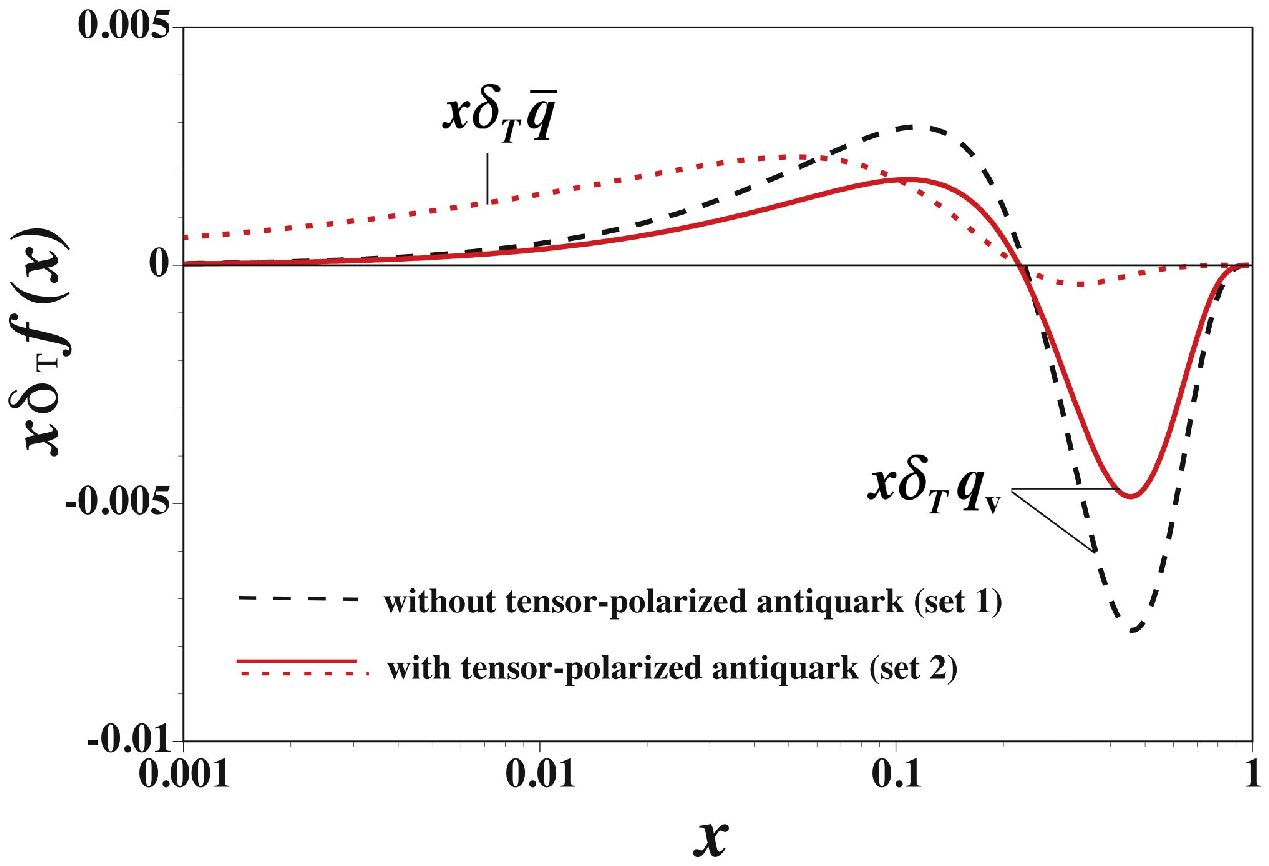}
  $ \ \ $
\label{fig:b1-motivation}
\begin{minipage}[c]{0.47\textwidth}
\setlength{\baselineskip}{10pt}
\vspace{0.5cm}
\noindent
{\footnotesize 
 {\bf FIGURE 7.} 
 Two analysis results are shown with the HERMES $b_1$ data
 \cite{sk-b1-analysis}.}
\end{minipage}
\hspace{0.55cm}
\begin{minipage}[c]{0.47\textwidth}
\setlength{\baselineskip}{10pt}
\vspace{0.5cm}
{\footnotesize 
 {\bf FIGURE 8.} Obtained tensor-polarized distributions
                 \cite{sk-b1-analysis}.
 }
\end{minipage}
\end{minipage}
\end{figure}

In Fig.7, it is obvious that much better measurements are 
needed to investigate more details.
Because there is a proposal to measure $b_1$ at JLab \cite{jlab-b1},
much details of the tensor structure will be investigated in 
a few years.
On the other hand, an appropriate future experiment to probe
the antiquark tensor polarization is a Drell-Yan experiment,
for example, by proton-deuteron Drell-Yan with a polarized
deuteron target \cite{pd-drell-yan}. It could be possible
at hadron facilities such as
J-PARC (Japan Proton Accelerator Research Complex) \cite{j-parc} and
GSI-FAIR (Gesellschaft f\"ur Schwerionenforschung -Facility for 
Antiproton and Ion Research).
Hadron-tensor studies become a new era in terms of
quark and gluon degrees of freedom!

\vfill\eject

\section{CDF dijet anomaly within standard model}\label{cdf}
\vspace{-0.10cm}

The CDF anomaly \cite{cdf-w2j-2011} was observed
in the dijet-mass distribution in $p \bar p$ collisions
by observing events with two energetic jets,
one high-$p_T$ electron or muon, and missing $E_T$.
However, another experimental group $D0$ did not observe
the same phenomenon \cite{d0}.
Here, we investigate a possibility to explain it 
within the standard model as effects of the PDFs. 
A fraction of the proton or antiproton beam energy 980 GeV
(=1.96 TeV /2) needs to be transferred to one of the dijets
with the energy about 70 GeV (=140 GeV /2), so that
PDFs of the $x=0.1$ region affect the CDF result.
Since light-quark ($u$ and $d$) distributions are well determined
at $x \sim 0.1$, we focus our attention to the strange-quark
distribution $s(x)$. In particular, the Bjorken variable-$x$
dependence of $s(x)$ is not known.
The 2nd moment of $s(x)$ is determined by 
the neutrino-induced dimuon measurements in comparison
with the 2nd moment of the light-antiquark distributions.
Then, a similar or same $x$-dependent functional form is
assumed for $s(x)$ as $[\bar u(x)+\bar d(x)]/2$.
It is typically illustrated by the HERMES experiment 
\cite{hermes-sx}, which indicates a much softer $s(x)$
distribution as shown in Fig.9.
The CDF dijet anomaly could be related to
such an issue of the strange-quark distribution. 

\vspace{0.0cm}
\begin{figure}[b]
\begin{minipage}[c]{\textwidth}
   \vspace{0.0cm}
   \includegraphics[width=0.48\textwidth]{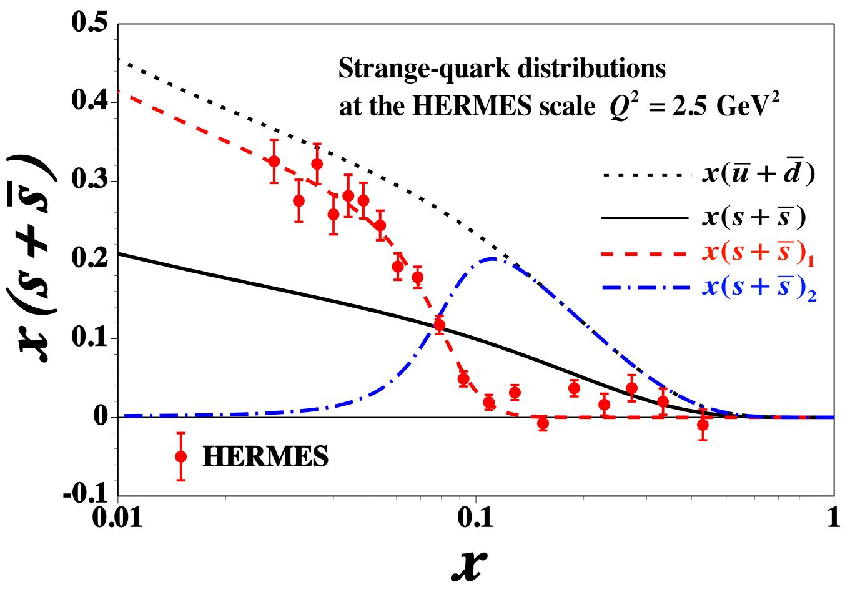}
   \hspace{0.3cm}
   \includegraphics[width=0.47\textwidth]{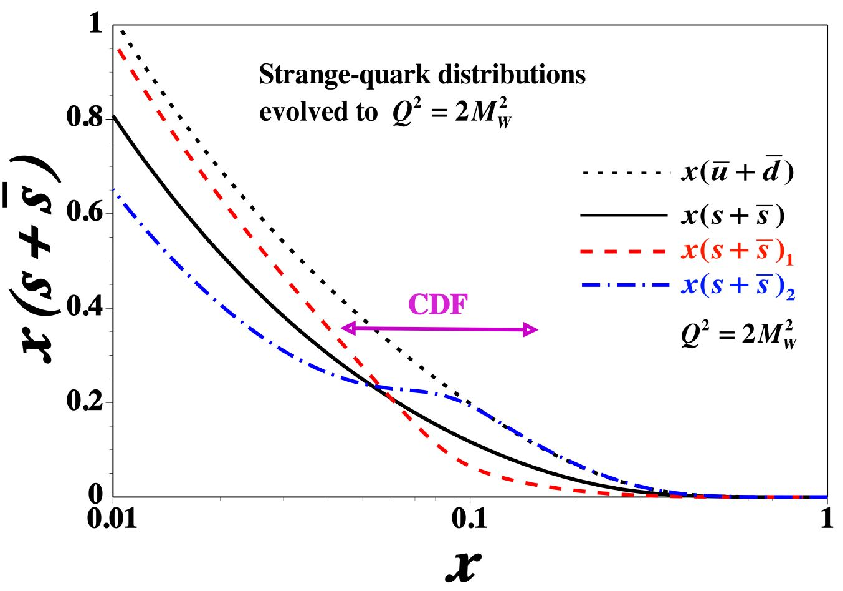}
  $ \ \ $
\label{fig:cdf-ano}
\begin{minipage}[c]{0.48\textwidth}
\setlength{\baselineskip}{10pt}
\vspace{0.4cm}
\noindent
{\footnotesize 
 {\bf FIGURE 9.} 
 Strange-quark distribution by HERMES \cite{hermes-sx}, 
 CTEQ6L1 distribution \cite{cteq6l1}, 
 and a possible hard strange-quark distribution \cite{kek-2011}.}
\end{minipage}
\hspace{0.55cm}
\begin{minipage}[c]{0.49\textwidth}
\setlength{\baselineskip}{10pt}
\vspace{0.3cm}
{\footnotesize \hspace{0.10cm}
 {\bf FIGURE 10.} The strange-quark distributions
 
\noindent
\hspace{0.1cm}
  in Fig.9 are evolved to the scale $Q^2=2 M_W^2$ 

\noindent
\hspace{0.1cm} 
  in the CDF experiment \cite{kek-2011}.
 }
\end{minipage}
\end{minipage}
\end{figure}

The strange-quark distribution is roughly 40\% of
the light antiquark distribution$[\bar u(x)+\bar d(x)]/2$
as shown by the solid curve in Fig.9. 
However, recent HERMES experiment indicated
the much softer distribution as it becomes
close to $x (\bar u+\bar d)$ at small $x$ and it almost vanishes
at $x>0.1$. The HERMES measurement was done by semi-inclusive
kaon production in charged-lepton deep inelastic scattering.
It should be noticed that kaon fragmentation functions have
large uncertainties as typically shown in Ref. \cite{hkns07},
so that the HERMES determination of $s(x)$ may not be accurate.
Therefore, we take a hard strange-quark distribution shown
by the dot-dashed curve as one of trial functions
in estimating PDF effect on the dijet
cross section in addition to the CTEQ6L1
and HERMES-like soft distribution.
Like the intrinsic charm distribution at large $x$ \cite{intrinsic-c},
there could be intrinsic strange distribution 
\cite{flavor, strange-large-x}
which gives rise to the excess of $s(x)$ at large $x$.
The average scale of the HERMES experiment is $Q^2$=2.5 GeV$^2$,
so that the distributions should be evolved to a hard scale
in the CDF experiment.
The strange-quark distributions $s+\bar s$, $(s+\bar s)_1$, and 
$(s+\bar s)_2$ in Fig.9 are evolved to the typical hard
scale $Q^2=2 M_W^2$ in the CDF dijet measurement
by the $Q^2$ evolution code of Ref. \cite{Q2-code}.
The evolved distributions are shown in Fig.10 together with
$x(\bar u+\bar d)$ at $Q^2=2 M_W^2$.
If the distributions are evolved to such a large scale,
the differences between the three distributions are not 
as large as the ones in Fig.9 because strange quarks 
are copiously produced by the $Q^2$ evolution.

\begin{wrapfigure}{r}{0.50\textwidth}
   \vspace{0.3cm}
   \begin{center}
       \epsfig{file=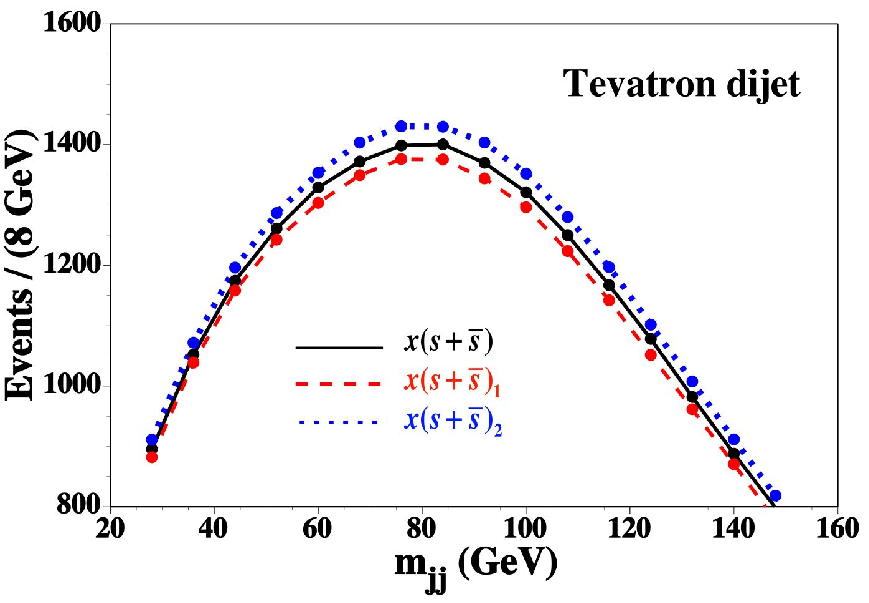,width=0.45\textwidth} \\
   \end{center}
   \vspace{-0.2cm}
{\footnotesize \hspace{0.30cm}
 {\bf FIGURE 11.} Effects of $s(x)$ on dijet events
 \cite{kek-2011}.
 }
\vspace{0.2cm}
\end{wrapfigure}

There are various processes which contribute to the dijet events.
We calculated the processes $W$+dijet, $Z$+dijet, top,
$WW$, and $WZ$ by the event generator
GR@PPA (GRace At Proton-Proton/Antiproton collisions) \cite{gr@ppa}.
By supplying kinematical conditions and a reaction process,
the GR@PPA automatically calculates possible processes and
their contributions to a cross section.
Effects of $s(x)$ on the dijet cross section are shown in Fig.11.
In our work, the partonic cross sections are calculated, and
subsequent parton shower and final fragmentations 
into hadrons are not included. They could be calculated, 
for example, by using the event generator PYTHIA.
It is the purpose of this work to show gross properties
of strange-quark effects simply by calculating the hard
process part.
Furthermore, since the detector acceptance information
is not available for public, the overall magnitude of the events
cannot be compared with the CDF measurements. We may
look at overall shapes of the events.

According to the results in Fig.11,
the dijet-mass distribution increases if the hard distribution
$(s+\bar s)_2$ is taken, whereas it decreases for the soft
one $(s+\bar s)_1$. We also notice the shape of the dijet-mass
distribution changes depending on the strange-quark distribution.
Since the CDF anomaly was observed in the shoulder region of
$m_{jj}\approx$140 GeV, a slight change of the dijet-mass
distribution could explain at least partially the anomalous excess
if it is not a strong peak \cite{kek-2011}.
We also calculated effects of the strange-quark modifications
on the dijet cross section at LHC. We found sizable effects
in the LHC kinematics; however, the results are sensitive
to a different $x$ region ($x \sim 0.02$) because of
the larger center-of-mass energy of 14 TeV. Therefore,
the $x$ dependence of the strange-quark distribution
should be constrained by future LHC measurements
together with Tevatron data.

In our studies, the possible variations of $s(x)$ are
investigated and their effects on the CDF dijet cross sections
are calculated. We found that the effects are sizable in
$W$+dijet and $Z$+dijet processes and that they are very
small in other processes (top, $WW$, $WZ$).
It is important to consider such PDF effects within 
the standard model for testing whether the CDF anomaly 
actually indicates new physics beyond the standard model. 
Further theoretical and experimental investigations
are needed including physics mechanisms of creating
the strange-quark distribution other than the one
produced by the obvious $Q^2$ evolution.

\section{Acknowledgements}
\vspace{-0.2cm}
This work was partially supported by a Grant-in-Aid for Scientific 
Research on Priority Areas ``Elucidation of New Hadrons with a Variety 
of Flavors (E01: 21105006)" from the ministry of Education, Culture, 
Sports, Science and Technology of Japan.  



\end{document}